\def\isarxiv{1} 

\ifdefined\isarxiv
\documentclass[11pt]{article}

\usepackage[numbers]{natbib}

\else
\documentclass{article}
\usepackage{neurips_2022}
\fi

\usepackage{amsmath}
\usepackage{amsthm}
\usepackage{amssymb}
\usepackage{algorithm}
\usepackage{subfig}
\usepackage{algpseudocode}
\usepackage{graphicx}
\usepackage{grffile}
\usepackage{wrapfig,epsfig}
\usepackage{url}
\usepackage{xcolor}
\usepackage{epstopdf}
\usepackage{bbm}
\usepackage{dsfont}

\allowdisplaybreaks

\ifdefined\isarxiv

\usepackage{tikz}
\usepackage{hyperref}  
\hypersetup{colorlinks=true,citecolor=blue,linkcolor=blue} 
\usetikzlibrary{arrows}
\usepackage[margin=1in]{geometry}

\else

\usepackage{microtype}
\usepackage{hyperref}
\definecolor{mydarkblue}{rgb}{0,0.08,0.45}
\hypersetup{colorlinks=true, citecolor=mydarkblue,linkcolor=mydarkblue}

\fi

\newtheorem{theorem}{Theorem}[section]
\newtheorem{lemma}[theorem]{Lemma}
\newtheorem{definition}[theorem]{Definition}

\newtheorem{fact}[theorem]{Fact}

\newcommand{\wh}{\widehat}
\newcommand{\wt}{\widetilde}

\newcommand{\R}{\mathbb{R}}

\newcommand{\Tmat}{{\cal T}_{\mathrm{mat}}}
\newcommand{\eps}{\epsilon}

\DeclareMathOperator{\poly}{poly}
\DeclareMathOperator{\A}{\mathsf{A}}
\DeclareMathOperator{\B}{\mathsf{B}}

\DeclareMathOperator{\vect}{vec}
\DeclareMathOperator{\tr}{tr}

\DeclareMathOperator{\new}{new}

\makeatletter
\newcommand*{\RN}[1]{\expandafter\@slowromancap\romannumeral #1@}
\makeatother

\usepackage{lineno}

\begin{document}

\ifdefined\isarxiv

\date{\empty}

\title{Streaming Semidefinite Programs: $O(\sqrt{n})$ Passes, Small Space and Fast Runtime}
\author{
Zhao Song\thanks{\texttt{zsong@adobe.com}. Adobe Research.}
\and
Mingquan Ye\thanks{\texttt{mye9@uic.edu}. University of Illinois Chicago.}
\and
Lichen Zhang\thanks{\texttt{lichenz@mit.edu}. Massachusetts Institute of Technology.}
}

\else

\title{Intern Project} 
\maketitle 
\fi

\ifdefined\isarxiv
\begin{titlepage}
  \maketitle
  \begin{abstract}

We study the problem of solving semidefinite programs (SDP) in the streaming model. Specifically, $m$ constraint matrices and a target matrix $C$, all of size $n\times n$ together with a vector $b\in \R^m$ are streamed to us one-by-one. The goal is to find a matrix $X\in \R^{n\times n}$ such that $\langle C, X\rangle$ is maximized, subject to $\langle A_i, X\rangle=b_i$ for all $i\in [m]$ and $X\succeq 0$. Previous algorithmic studies of SDP primarily focus on \emph{time-efficiency}~\cite{lsw15,jkl+20,jlsw20,hjs+22}, and all of them require a prohibitively large $\Omega(mn^2)$ space in order to store \emph{all the constraints}. Such space consumption is necessary for fast algorithms as it is the size of the input. In this work, we design an interior point method (IPM) that uses $\widetilde O(m^2+n^2)$ space, which is strictly sublinear in the regime $n\gg m$. Our algorithm takes $O(\sqrt n\log(1/\epsilon))$ passes, which is standard for IPM. Moreover, when $m$ is much smaller than $n$, our algorithm also matches the time complexity of the state-of-the-art SDP solvers~\cite{jkl+20,hjs+22}. To achieve such a sublinear space bound, we design a novel sketching method that enables one to compute a spectral approximation to the Hessian matrix in $O(m^2)$ space. To the best of our knowledge, this is the first method that successfully applies sketching technique to improve SDP algorithm in terms of space (also time).

  \end{abstract}
  \thispagestyle{empty}
\end{titlepage}

{
}
\newpage

\else

\begin{abstract}

We study the problem of solving semidefinite programs (SDP) in the streaming model. Specifically, $m$ constraint matrices and a target matrix $C$, all of size $n\times n$ together with a vector $b\in \R^m$ are streamed to us one-by-one. The goal is to find a matrix $X\in \R^{n\times n}$ such that $\langle C, X\rangle$ is maximized, subject to $\langle A_i, X\rangle=b_i$ for all $i\in [m]$ and $X\succeq 0$. Previous algorithmic studies of SDP primarily focus on \emph{time-efficiency}~\cite{lsw15,jkl+20,jlsw20,hjs+22}, and all of them require a prohibitively large $\Omega(mn^2)$ space in order to store \emph{all the constraints}. Such space consumption is necessary for fast algorithms as it is the size of the input. In this work, we design an interior point method (IPM) that uses $\widetilde O(m^2+n^2)$ space, which is strictly sublinear in the regime $n\gg m$. Our algorithm takes $O(\sqrt n\log(1/\epsilon))$ passes, which is standard for IPM. Moreover, when $m$ is much smaller than $n$, our algorithm also matches the time complexity of the state-of-the-art SDP solvers~\cite{jkl+20,hjs+22}. To achieve such a sublinear space bound, we design a novel sketching method that enables one to compute a spectral approximation to the Hessian matrix in $O(m^2)$ space. To the best of our knowledge, this is the first method that successfully applies sketching technique to improve SDP algorithm in terms of space (also time).

\end{abstract}

\fi

\section{Introduction}

Semidefinite programming (SDP) is one of the central problems in convex optimization, machine learning, and theoretical computer science. It serves as a powerful tool for obtaining approximation algorithms for classic combinatorial optimization problems, such as max-cut~\cite{gw94}, coloring $3$-colorable graphs \cite{kms94}, sparsest cut \cite{arv09}. In recent years, it also fuels the development for fast linear algebraic problems, such as spectral sparsification~\cite{ls17}, algorithmic discrepancy and rounding~\cite{bdg16,bg17,b19,bdg19,dgl19,hrs21,jrt23}, and terminal embeddings~\cite{cn21}. Optimizing over sum-of-squares polynomials can also be readily reduced to solving an SDP, and efficient algorithms are derived in a similar favor~\cite{jnw22}. SDP also functions as a primitive for machine learning algorithms with provable guarantees, such as adversarial learning~\cite{rsl18}, learning structured distribution~\cite{clm20}, sparse principal component analysis~\cite{aw08,dejl07}, robust statistics~\cite{dkklms16,dhl19,jlt20}, and matrix completion~\cite{ct10,r11,cr12}.

Formally, we define the SDP problem as follows. 
\begin{definition}[SDP instance]\label{def:general_sdp}
Given $m$ symmetric matrices $A_1, A_2, \cdots , A_m \in \R^{n \times n}$, a vector $b \in \R^m$, and a symmetric matrix $C \in \R^{n \times n}$, the goal is to optimize
\begin{align*}
    \max_{X\in \R^{n\times n}} & ~ \langle C , X \rangle \\
    \mathrm{s.t.~} & ~ \langle A_i, X \rangle = b_i,\ \forall i \in [m], \\
    & ~ X \succeq 0.
\end{align*}   
\end{definition}

Streaming algorithms are important because they enable efficient processing of data that arrives in a continuous and rapid stream, without the need to store entire dataset at once \cite{m14}. This is of particularly interest in the case of SDP, as the input size of an SDP is $\Theta(mn^2)$ and storing all data would potentially consume prohibitively large space when both $m$ and $n$ are large. Several prior works have studied the space and pass complexity of linear programming (LP)~\cite{akz19,lsz+23,bks23,bs23}, a special case of SDP.

In this paper, we investigate the complexity of solving SDP in the streaming model, where the constraint matrices $A_i\in \R^{n\times n}$ together with the objective matrix $C\in \R^{n\times n}$ and target vector $b\in \R^m$ are streamed to us one-by-one. From an algorithmic perspective, two types of algorithms for solving SDP have been widely studied both in theory and practice: when one only requires to solve the program in low accuracy (i.e., the complexity of the algorithm scales polynomially with respect to $1/\epsilon$), first-order method based on multiplicative weights update (MWU) can be utilized~\cite{ahk12,jy11,pt12,jy12,alo16}. In most of practical applications, high-accuracy solvers whose complexity scales with $\log(1/\epsilon)$ are oftentimes preferred. These methods are either based on interior pint method (IPM) or cutting plane method (CPM)~\cite{s77,yn76,k80,kte88,nn92,nn94,nn89,v89,a00,km03,bv02,lsw15,jlsw20,jkl+20,hjs+22,gs22}, and from a runtime perspective, current state-of-the-art algorithms all exploit the structure of IPM~\cite{jkl+20,hjs+22}. We focus on obtaining a space- and pass-efficient high-accuracy streaming algorithm as they are much more applicable than their low-accuracy counterparts. We note that prior time-efficient IPM algorithms for SDP all require to maintain the size $m\times n^2$ constraint matrices, which would already consume $\Omega(mn^2)$ space. Therefore, it is natural to ask that 
\begin{center}
    {\it Is it possible to solve SDP in sublinear space of the input size and in a pass-efficient manner?}
\end{center}

In this work, we provide an affirmative answer to the above question. Our key technical innovation is a tensor-based sketch that computes a compact representation of the Hessian matrix. To the best of our knowledge, this is the first successful deployment of sketching to SDP problem. Perhaps what's more surprising is that our algorithm is also \emph{time-efficient}: whenever $m\leq n^{0.25}$, a popular parameter regime for many SDP problems, the runtime of our algorithm matches the state-of-the-art SDP solvers~\cite{jkl+20,hjs+22}.

\subsection{Our Results}

 We state our main result as follows. 
\begin{theorem}[Main result, informal version of Theorem~\ref{thm:main_ipm_sdp_formal}]\label{thm:main_ipm_sdp}
Given an SDP instance (Definition~\ref{def:general_sdp}), there is an $O( \sqrt{n} \log(n/\epsilon))$-pass streaming algorithm that uses $\wt{O}(m^2 + n^2)$ space\footnote{We use $\wt O(\cdot)$ to suppress polylogarithmic factors in $m$ and $n$.} to solve SDP up to $\epsilon$ accuracy with probability $1-1/\poly(n)$. Moreover, the algorithm has a runtime complexity of $\wt O(\sqrt n (m^{\omega-1}n^2+n^\omega)\log(1/\epsilon))$, where $\omega$ is the exponent of fast matrix multiplication.
\end{theorem}
Currently $\omega \approx 2.37$ \cite{w12,lg14,aw21,dwz23,lg23,wxxz23}.

One of the popular parameter regimes is that the number of constraints $m$ is much smaller than the dimension $n$ ($m\ll n$). Many combinatorial problems and numerical linear algebraic problems admit such a setting, including the SDP formulation of support vector machines (SVM)~\cite{j06,cl11,gsz23}. This setting also naturally encapsulates the study of better iteration complexity of SDP, as most SDP solvers attempt to operate on the dual formulation of the problem, formulated by
\begin{align*}
    \min_{y\in \R^m} & ~ b^\top y \\
    \text{s.t.} & ~ \sum_{i=1}^m y_iA_i \succeq C,
\end{align*}
where the constraint set of the dual is an $m$-dimensional spectrahedron. The iteration complexity of IPM depends on the complexity of a self-concordant barrier function~\cite{nn94}, and given an $m$-dimensional convex body, it is natural to develop a barrier function that has complexity nearly linear in the dimension and can be computed efficiently. In the case of linear programming, Lee and Sidford~\cite{ls14,ls19} show that for an $n$-dimensional polytope, it is possible to compute a self-concordant barrier with complexity $\wt O(n)$ and its corresponding first- and second-order information in polynomial time. As the log-barrier for SDP has complexity $n$, it is natural to search for a better barrier function with complexity nearly linear in $m$ for $m\ll n$ and its first- and second-order information can be quickly computed\footnote{If one only looks for a barrier function with complexity linear in the dimension, the universal barrier function is sufficient~\cite{nn94,ly21}. However, the universal barrier is not known to be polynomial time computable.}. 

Under this regime, the space bound we obtain is essentially $\wt O(n^2)$, meaning that we \emph{merely store a constant number of constraint matrices.} This is surprising, as computing the Hessian matrix with respect to the log-barrier would either require one to explicitly write down all constraint matrices in $\Theta(mn^2)$ space, or use $O(n^2)$ space to compute a single entry of the Hessian but spend $m$ passes over the data. We present a novel approach to sketch the normalized constraint matrices and generate a spectral approximation of the Hessian matrix.

From the runtime perspective, when $m$ is in the interval $[n^{0.1},n^{0.27}]$, the state-of-the-art SDP solvers have the complexity of $\wt O(n^{\omega+0.5}\log(1/\epsilon))$~\cite{jkl+20}, and our algorithm runs in time $\wt O((n^{\omega+0.5}+m^{\omega-1}n^{2.5})\log(1/\epsilon))$. The $m^{\omega-1}n^{2.5}$ term is subsumed by the $n^{\omega+0.5}$ term, thus our algorithm has its runtime \emph{matches} the state-of-the-art solver. Notably, our algorithm is much simpler than that of~\cite{jkl+20,hjs+22}, where both of them rely on sophisticated low-rank maintenance data structures and potential analyses. In contrast, we show that whenever $m$ is much smaller than $n$, we can quickly generate a spectral approximation of the Hessian matrix and adapt the standard Newton IPM on the dual program. This opens up the gate for \emph{practical implementation} of our sketching scheme due to its simplicity and the success in practice.

\subsection{Technique Overview}

Given $A_1,\ldots,A_m\in \R^{n\times n}$, we use matrix ${\sf A}=\begin{bmatrix}
    \vect(A_1)^\top \\
    \vect(A_2)^\top \\
    \vdots \\
    \vect(A_m)^\top
\end{bmatrix}\in \R^{m\times n^2}$ to denote the batched constraint matrix, where $\vect(\cdot)$ is the vectorization operation that flattens a matrix of size $n\times n$ to an $n^2$-dimensional vector. Let $S(y):=C-\sum_{i=1}^m y_iA_i\in \R^{n\times n}$ denote the slack matrix, and when $y$ is clear from context, we just use $S$. The major computation bottleneck for interior point method with log-barrier is to compute the Hessian and its inverse, where the Hessian matrix is $m\times m$ and can be compactly written as $H(y)={\sf A} (S^{-1}\otimes S^{-1}){\sf A}^\top$ with $\otimes$ being the Kronecker product. Although computing the slack matrix given $y$ only requires $O(n^2)$ space and one pass over the data, explicitly expressing the Hessian would mandate $\Theta(mn^2)$ space to store the matrix ${\sf A}$. Alternatively, one can use the property of vectorization and Kronecker product to show that each entry of $H(y)$ is equivalent to
\begin{align*}
    H(y)_{i,j} = & ~ \tr[S^{-1/2}A_iS^{-1/2}\cdot S^{-1/2}A_jS^{-1/2}]. 
\end{align*}
Hence, writing down a single entry only needs $O(n^2)$ space and it is tempting to slowly fill in the Hessian by performing this operation in place. Unfortunately, as we cannot afford to store all of the constraint matrices, each time we compute an entry and have to query corresponding constraint matrices leading to a total of $\Theta(m)$ passes. Compounded with the $O(\sqrt n\log(1/\epsilon))$ iterations of IPM, this would imply an overall $O(m\sqrt n\log(1/\epsilon))$ passes, which is far from optimal whenever $m$ is relatively large. 

A popular approach in streaming to improve the space efficiency is via linear sketches~\cite{ams96,ccf02} which are a distribution of random matrices that preserve certain statistics of the data stream with a small number of rows. In our setting, we would like to generate a \emph{spectral approximation} of the Hessian matrix in small space, and a natural candidate is the linear sketches satisfying the \emph{subspace embedding} property~\cite{s06}\footnote{We remark that subspace embedding is one of key property being used in giving the state-of-the-art result for linear regression and low-rank approximation \cite{cw13,nn13}.}. Roughly speaking, given a tall and skinny matrix $B\in \R^{n\times d}$, a sketching matrix $R$ with $(\epsilon,\delta)$-subspace embedding property would satisfy that with probability at least $1-\delta$, for any vector $x\in \R^d$, $\|RBx\|_2^2=(1\pm\epsilon)\cdot \|Bx\|_2^2$. Moreover, $R$ has much fewer rows than $n$. This means that the matrix $B^\top R^\top RB$ is a good spectral approximation to $B^\top B$ with high probability. Unfortunately, even a sketch matrix $R$ is explicitly given to us, it is unclear how can we directly apply to the Hessian, as our Gram matrix is in the form of ${\sf A}(S^{-1}\otimes S^{-1}){\sf A}^\top$. Moreover, the size of the sketching matrix is $\wt O(mn^2)$, meaning that explicit representation of the sketching matrix cannot be stored. While many space-efficient sketches only require a few bits for random seeds~\cite{kn14}, applying them to a target matrix is sometimes much harder when an implicit Hessian matrix is streamed to us.

Our solution is inspired by recent developments of applying sketching-based data structures to speed up iterative algorithms~\cite{lsz19,jswz21,sy21,bpsw21,szz21,sxz22,gs22,qszz23,syyz23,syyz23_weight,gsz23,gsyz23,z22}. In these applications, sketching matrices are usually tailored towards particular iterative templates, so that they can accommodate for different robust guarantees posed by these algorithms. The key structure to leverage here is the Kronecker formulation of the Hessian: using the mixed product property, we can rewrite the Hessian as $H(y)={\sf A}(S^{-1/2}\otimes S^{-1/2})(S^{-1/2}\otimes S^{-1/2}) {\sf A}^\top$, therefore we only need to consider sketching $(S^{-1/2}\otimes S^{-1/2}){\sf A}^\top$. If one can manage to devise a sketching scheme that handles the Kronecker product $S^{-1/2}\otimes S^{-1/2}$ without explicitly forming them, we can then use this sketch of the Kronecker product to multiply with each row of ${\sf A}$. This enables us to compute a lossy spectral approximation in $\wt O(m^2)$ space and $O(1)$ passes over the data. From a dimensionality reduction perspective, many sketches have been particularly tuned for inputs with tensor structure, so that the large $n^2$ dimensions can be reduced, the subspace induced by these tensors can be preserved and these sketches can be quickly applied. For example, count sketch matrix can be combined with Fast Fourier Transform (FFT) to quickly approximate the tensor product of two vectors so that both space and time efficiency can be improved~\cite{pp13}. This so-called {\sf TensorSketch} is later proved to have the subspace embedding property~\cite{anw14} and utilized for Kronecker product regression~\cite{dssw18,djssw19,rsz22} and tensor low-rank approximation \cite{swz19_soda}. Another kind of sketches for tensor product is the {\sf TensorSRHT} matrix~\cite{akk+20}, which provides subspace embedding guarantee with high probability. {\sf TensorSRHT} has wide range of applications for sketching polynomial kernels and neural tangent kernels~\cite{swyz21,szz21,zha+21,wz22}. Our algorithm utilizes the {\sf TensorSRHT} matrix since its row count depends polylogarithmically on $1/\delta$, the reciprocal of the success probability\footnote{An alternative sketching matrix to use is the {\sf TensorSparse} proposed by \cite{sxz22}. Using that construction will get the same result as {\sf TensorSRHT} up to log factors.}. However, it still remains unclear how one can even apply the {\sf TensorSRHT} matrix to $(S^{-1/2}\otimes S^{-1/2}){\sf A}^\top$.

Let us examine columns of the matrix $(S^{-1/2}\otimes S^{-1/2}){\sf A}^\top$, which are $n^2$-dimensional vectors in the form of $(S^{-1/2}\otimes S^{-1/2})\vect(A_i)$. The {\sf TensorSRHT} matrix can be written as $\Pi:=P(HD_1\otimes HD_2)$, where $D_1, D_2\in \R^{n\times n}$ are diagonal matrices with diagonal entries being random Rademacher, $H$ is the $n^2\times n^2$ Hadamard matrix, and $P$ is a row sampling matrix that samples $s=\Theta(\epsilon^{-2}m\log^3(nm/(\epsilon\delta)))$ rows with a normalization factor $\frac{1}{\sqrt s}$. To apply $\Pi$ to $(S^{-1/2}\otimes S^{-1/2}){\sf A}^\top$, we first use the mixed product property to compute $(HD_1\otimes HD_2)\cdot (S^{-1/2}\otimes S^{-1/2})=HD_1S^{-1/2}\otimes HD_2S^{-1/2}$, then note that multiplying with $\vect(A_i)$ can be converted as follows:
\begin{align*}
    (HD_1S^{-1/2}\otimes HD_2S^{-1/2})\vect(A_i) = & ~ \vect(HD_2S^{-1/2}A_iS^{-1/2}D_1 H^\top).
\end{align*}

Given this $n^2$-dimensional vector, we can then perform subsequently row sampling simply as sampling coordinates. The resulting vector is of dimension $m$, as we repeat this procedure for all of $m$ columns of ${\sf A}^\top$, yielding an algorithm that uses $O(sm+n^2)$ space and $O(1)$ passes to generate a spectral approximation of the Hessian. We note that this procedure can also be made \emph{time-efficient}: the process of computing $HD_1S^{-1/2}$ and $HD_2S^{-1/2}$ takes $O(n^2\log n)$ time by leveraging fast Hadamard transform, and to perform the coordinate sampling, we avoid forming the Kronecker product explicitly. We interpret $P$ as sampling $s$ entries from the matrix $HD_2S^{-1/2}A_i S^{-1/2}D_1H^\top$. Let $(i_1,j_1),\ldots,(i_s,j_s)$ denote these sampled entries, and we can construct two matrices $X, Y\in \R^{s\times n}$ satisfying that the $k$-th row of $X$ is $(HD_2S^{-1/2})_{i_k,*}$ and the $k$-th row of $Y$ is $(HD_1S^{-1/2})_{j_k,*}$, then the sampled entries of the target matrix can be read from the $(i_k,j_k)$-th entry of the product $XA_iY^\top\in \R^{s\times s}$. The time for constructing $X$ and $Y$ is $O(sn)$, and computing the compact target matrix is $\Tmat(s,n,n)$, where $\Tmat(a,b,c)$ denotes the complexity of multiplying an $a\times b$ matrix with a $b\times c$ matrix. Repeating this procedure for $m$ columns, this amounts to a total time of 
\begin{align*}
    m\cdot \Tmat(s,n,n).
\end{align*}
If $\epsilon=O(1)$ and $\delta=1/\poly(n)$, then this translates to a total runtime of $\wt O(n^2m^{\omega-1})$, while explicitly forming the Hessian matrix would take $O(mn^\omega)$ time. Our algorithm is superior whenever $m\ll n$.

Our sketching scheme also implies a simple, log-barrier based SDP algorithm with a total runtime of 
\begin{align*}
    \wt O((n^{2.5}m^{\omega-1}+n^{\omega+0.5})\log(1/\epsilon)). 
\end{align*}
For the parameter regime $m\in [n^{0.1}, n^{\frac{\omega-2}{\omega-1}}]$, we have that $n^{2.5}m^{\omega-1}\leq n^{\omega+0.5}$ and thus match the currently best SDP solver due to Jiang, Kathuria, Lee, Padmanabhan, and Song~\cite{jkl+20}\footnote{We require $m\geq n^{0.1}$ because the hybird barrier-based algorithm due to Huang, Jiang, Song, Tao, and Zhang~\cite{hjs+22} is faster when $m\leq n^{0.1}$.}.

\subsection{Open Problems}

In this paper, we study the problem of solving SDP in the streaming model, where one aims to improve space usage and the number of passes over the data. We present an algorithm that uses $\wt O(m^2+n^2)$ space and $O(\sqrt{n}\log(1/\epsilon))$ passes. Moreover, our algorithm runs in time $\wt O(\sqrt{n}(m^{\omega-1}n^2+n^\omega)\log(1/\epsilon))$ which matches the state-of-the-art SDP solver runtime when $m\ll n$. Our algorithm relies on a novel application of the {\sf TensorSRHT} matrix that generates a quick spectral approximation to the Hessian matrix in small space. We leave several open problems to be solved.

\paragraph{Improve the Space Bound to $\wt O(m+n^2)$.} In the regime where $m\ll n$, our algorithm achieves a nearly-optimal space bound, as the $\Omega(n^2)$ space seems unavoidable at least for storing a constant number of the constraint matrices and constructing the slack matrix. In the regime where $m\gg n$, our space bound becomes $\wt O(m^2)$, i.e., the space required to store the Hessian matrix of the IPM. On the other hand, the constraint set we are optimizing is the $m$-dimensional spectrahedron, therefore \emph{is it possible to use $\wt O(m+n^2)$ space to solve SDP in the streaming model?} This means that we are only allowed to store at most polylogarithmically many constraint matrices and dual variables, and we do not have the space budget to explicitly construct and store the Hessian (even a spectral approximation of the Hessian).

\paragraph{IPM Beyond $O(\sqrt n)$ Passes.} The $O(\sqrt n\log(1/\epsilon))$-pass bound follows from implementing log-barrier based IPM, since the IPM converges in $O(\sqrt n\log(1/\epsilon))$ iterations and in each iteration, we make sure to go through $O(1)$ passes of the data to obtain the desired pass bound. This is relatively unfavourable when $n$ is large, and one can utilize hybrid barrier for a better iteration complexity~\cite{hjs+22}. However, computing the Hessian of hybrid barrier would need to compute much more sophisticated numerical measurements of the constraints, therefore it is unclear whether one can still maintain the $\wt O(m^2+n^2)$ space bound we obtained.

\paragraph{Low-Accuracy Regime.} Instead of using IPM-based second-order method, one can resort to low-accuracy first-order method, which would ideally improve the space usage. In particular, matrix multiplicative weights update (MMWU)~\cite{ak07,jy11,ahk12,alo16} is a popular first-order algorithm for solving packing and covering SDPs. It will be interesting to examine the space- and pass-complexity of these algorithms, and provide a deterministic algorithm for MMWU. We also notice that MMWU only works for packing and covering SDPs, so it is important to investigate the first-order streaming algorithm for SDP. It is also worth studying the space complexity of online covering SDP and its learning-augmented version, and trying to adapt them to the streaming settings~\cite{ekn16,glssz22}. 

\section{Preliminary}

Given two symmetric matrices $A,B$, we use $\langle A, B\rangle$ to denote their inner product, i.e., $\langle A, B \rangle = \tr[A^\top B]$. For a symmetric matrix $X$, let $e^X$ denote its matrix exponential, i.e., $e^X = \sum_{k=0}^{\infty} \frac{1}{k!} X^k$. For a matrix $A$, we use $\| A \|$ to represent its spectral norm. For a positive integer $n$, let $[n]$ represent the set $\{1, \cdots, n\}$. For a vector $x \in \R^m$, let $\| x \|_1$ and $\| x \|_2$ denote its $\ell_1$ and $\ell_2$ norm respectively. For a square matrix $A \in \R^{n \times n}$, we say $A$ is positive semidefinite ($A \succeq 0$) if for all $x^\top A x \geq 0$. We say $A$ is positive definite ($A \succ 0$) if for all non-zero $x$, we have $x^\top A x >0$. We use $\mathbb{S}_{\geq 0}^{n \times n}$ to denote the set of $n \times n$ positive semidefinite matrices. Given two matrices $A_1 \in \R^{n_1 \times d_1}$ and $A_2 \in \R^{n_2 \times d_2}$, we use $A_1 \otimes A_2 \in \R^{n_1 n_2 \times d_1 d_2}$ to denote the matrix where the $(i_1+(i_2-1) n_1, j_1+(j_2-1)d_1)$-th entry is $(A_1)_{i_1,j_1} (A_2)_{i_2,j_2}$ for all $i_1 \in [n_1], i_2 \in [n_2]$, $j_1 \in [d_1]$, and $j_2 \in [d_2]$.

\subsection{Correctness Guarantee via Robust IPM}

We state the correctness guarantee from the robust IPM framework developed in prior works.
\begin{lemma}[\cite{jkl+20,hjs+22}]\label{lem:correctness}
    Consider a semidefinite program with variable size $n \times n$ and $m$ constraints,  
    \begin{align*}
        \max\ &\langle C, X \rangle \\
        \mathrm{s.t.} \ & \langle A_i, X \rangle = b_i,\ \forall i \in [m], \\
             & X \succeq 0.
    \end{align*}
    Assume that any feasible solution $X \in \mathbb{S}_{\ge 0}^{n \times n}$ satisfies $\| X \|  \le R$, then for any error parameter $0 < \epsilon \le 0.01$, there is an interior point method that outputs a positive semidefinite matrix $\wh{X} \in \mathbb{S}^{n \times n}_{\ge 0}$ in $O(\sqrt{n} \log(n/\epsilon))$ iterations such that 
    \begin{align*}
        \langle C, \wh{X} \rangle &\ge \langle C, X^* \rangle - \epsilon \cdot \| C \| \cdot R, \\
        \sum_{i \in [m]} | \langle A_i, \wh{X} \rangle - b_i | &\le 4 n \epsilon ( R \sum_{i \in [m]} \| A_i \|_1 + \| b \|_1 ),
    \end{align*}
    where $X^*$ is the optimal solution to the semidefinite program, and $\| A_i \|_1$ is the Schatten $1$-norm of matrix $A_i$. 
\end{lemma}

\subsection{Tensor Subsampled Randomized Hadamard Transform}

Now we present a particular type of sketch for Kronecker product of matrices.

\begin{definition}[\textsf{TensorSRHT} \cite{akk+20,swyz21}]\label{def:tensor_srht}
The \textsf{TensorSRHT} $\Pi: \R^n \times \R^n \to \R^s$ is defined as $S := \frac{1}{\sqrt{s}} P \cdot (HD_1 \otimes HD_2)$, where each row of $P \in \{0, 1\}^{s \times n^2}$ contains only one $1$ at a random coordinate and one can view $P$ as a sampling matrix; $H$ is an $n \times n$ Hadamard matrix, and $D_1$, $D_2$ are two $n \times n$ independent diagonal matrices with diagonals that are each independently set to be a Rademacher random variable (uniform in $\{-1, 1\}$).  
\end{definition}

\begin{lemma}[\cite{akk+20}, see Lemma 2.12 in~\cite{swyz21} as an example]\label{lem:tensor_srht_implies_ose}
    Let $\Pi$ be a {\sf TensorSRHT} matrix defined in Definition~\ref{def:tensor_srht}. If $s=O(\epsilon^{-2}m \log^3(n m / (\epsilon\delta) ) )$, then for any orthonormal basis $U\in \R^{n^2\times m}$, with probability at least $1-\delta$, the singular values of $\Pi U$ lie in the range $[1-\epsilon, 1+\epsilon]$. 
\end{lemma}

Using fast Hadamard transform, the sketching matrix $\Pi$ can be applied to tensor product of two $n$-dimensional vectors in time $O(n\log n+s)$.

\section{Our Algorithm and Analysis}

\begin{algorithm}[!ht]\caption{A streaming algorithm for solving SDP with log-barrier.}
\label{alg:ours}
\begin{algorithmic}[1]
\Procedure{StreamingSDP}{$\mathsf{A} \in \R^{m \times n^2}, b \in \R^m, C \in \R^{n \times n}$, $\epsilon$} \Comment{Theorem~\ref{thm:main_ipm_sdp_formal}}
    \State $T \gets O( \sqrt{n} \log(n/\epsilon) )$
    \For{$t=1 \to T$} 
        \State $\eta^{\new} \gets \eta \cdot (1 + 1 / \sqrt{n})$
        \For{$j = 1 \to m$}
            \State $g_{\eta^{\new}}(y)_j \gets \eta^{\new} \cdot b_j - \tr[S^{-1} \cdot A_j]$ \Comment{$S \in \R^{n \times n}$, $A_j \in \R^{n \times n}$}
        \EndFor
        \State Compute $S^{-1/2}$
        \For{$j=1 \to m$}
            \State Compute $Q_i \gets \Pi (S^{-1/2} \otimes S^{-1/2}) \vect(A_i)$
        \EndFor
        \State Form $\wt{\B}$ by using $Q_1, \cdots, Q_m$ \Comment{$Q_i = \Pi (\B^\top)_{*,i}$}
        \State $\wt{H} \gets \wt{\B} \wt{\B}^\top$
        \State $\delta_y \gets - \wt{H}^{-1} g_{\eta^{\new}}(y)$
        \State $y^{\new} \gets y + \delta_y$
        \State $S^{\new} \gets \sum_{i \in [m]} ( y^{\new} )_i A_i - C$
        \State {\color{blue}/*Refresh memory*/}
        \State $\eta \gets \eta^{\new}$, $S \gets S^{\new}$, $y \gets y^{\new}$
    \EndFor 
\EndProcedure
\end{algorithmic}
\end{algorithm}

\subsection{Main Result}

\begin{theorem}[Main result, formal version of Theorem~\ref{thm:main_ipm_sdp}]\label{thm:main_ipm_sdp_formal}
Given an SDP instance (Definition~\ref{def:general_sdp}), there is an $O( \sqrt{n} \log(n/\epsilon))$-pass streaming algorithm that uses $O(n^2 + m^2 \poly\log(mn \log(n/\epsilon) ) )$ space to solve SDP up to $\eps$ accuracy and outputs a PSD matrix $X$ such that 
\begin{align*}
        \langle C, \wh{X} \rangle &\ge \langle C, X^* \rangle - \epsilon \cdot \| C \| \cdot R, \\
        \sum_{i \in [m]} | \langle A_i, \wh{X} \rangle - b_i | &\le 4 n \epsilon ( R \sum_{i \in [m]} \| A_i \|_1 + \| b \|_1 ),
    \end{align*}
where the optimal solution $X^*$ satisfies that $\| X^* \| \leq R$.

Moreover, our algorithm runs in time 
\begin{align*}
    O(\sqrt{n}(n^\omega+m^{\omega-1}n^2 \poly\log(mn))\log(1/\epsilon)).
\end{align*}
\end{theorem}
\begin{proof}

{\bf Proof of Space.} Through the course of the algorithm, we list the space for the following objects: 
\begin{itemize}
    \item $O(n^2)$ space for $S$, $S^{-1}$, and $S^{-1/2}$; 
    \item $\wt{O}(m^2)$ space for a sketch of version of $\B^\top$, denoted by $\wt{B}$; 
    \item $O(m^2)$ space for $\wt{H}$ and $\wt{H}^{-1}$; 
    \item $O(m)$ space for $y \in \R^m$; 
    \item $O(m)$ space for the gradient vector $g$.
\end{itemize}  

{\bf Proof of Passes.} The algorithm has $T$ iterations. In each iteration, we need three passes to scan $m$ matrices $A_1, \cdots, A_m$ one by one, thus the total number of passes is $O(T)$.

{\bf Proof of Runtime.} The most time-consuming steps of Algorithm~\ref{alg:ours} are computing $S^{-1/2}$, which would take $O(n^\omega)$ time, and computing the approximate matrix $\wt {\sf B}$, which would take time $O(m\cdot \Tmat(s,n,n))$ due to Lemma~\ref{lem:sketch_B}. Since we can choose the approximation factor to be $0.01$, we have $s=O(m\poly\log(mn))$ and this step takes time $O(m^{\omega-1}n^2)$, as desired.

{\bf Proof of Correctness.} Due to the robust IPM framework proposed in \cite{hjs+22}, it is  sufficient to give a constant approximation to Hessian, i.e.,
\begin{align*}
(1-0.01) H \preceq \wt{H} \preceq (1+0.01) H. 
\end{align*}
In the original framework of \cite{hjs+22}, they have many approximations, while in our streaming case, we only use approximate Hessian.

\end{proof}

\subsection{Rewrite Hessian}

\begin{fact}
If the following conditions hold
\begin{itemize}
    \item Let $S \in \R^{n \times n}$ denote a positive definite matrix.
    \item Let $\A \in \R^{m \times n^2}$ denote a matrix where each row is the vectorization of $A_i$.
    \item $H = \mathsf{A} (S^{-1} \otimes S^{-1}) \mathsf{A}$.
\end{itemize}
Then we have
\begin{align*}
    H_{i,j} = \tr[ S^{-1/2} A_i S^{-1/2} \cdot S^{-1/2} A_j S^{-1/2} ].
\end{align*}
\end{fact}

\begin{definition}
\label{def:sdp_B}
We define a matrix ${\sf B}\in \R^{m\times n^2}$ as 
\begin{align*}
{\sf B} := {\sf A} (S^{-1/2} \otimes S^{-1/2}).
\end{align*}
Then $H={\sf B}{\sf B}^\top$.
\end{definition}

We state a useful lemma for computing the matrix ${\sf B}$.

\begin{lemma}
Let matrix ${\sf B}\in \R^{m \times n^2}$ be defined as in Def.~\ref{def:sdp_B}. Then, the $i$-th row of ${\sf B}$ can be computed as $\vect(S(x)^{-1/2}A_iS(x)^{-1/2})$.
\end{lemma}

\begin{proof}
The proof relies on a simple fact of Kronecker product and vectorization:
\begin{align*}
\vect(S(x)^{-1/2}A_iS(x)^{-1/2}) = & ~ (S(x)^{-1/2}\otimes S(x)^{-1/2})\vect(A_i),
\end{align*}
which is the definition of the $i$-th row of ${\sf B}$.
\end{proof}

\subsection{Fast Hessian Approximation}

The following result provides an efficient embedding for $m \times n^2$ size matrix ${\sf B}$.

\begin{lemma}
\label{lem:sketch_B}
If the following conditions hold
\begin{itemize}
\item Let ${\sf B}\in \R^{ m \times n^2}$ be defined as in Def.~\ref{def:sdp_B}.
\item Let $\epsilon \in (0,1/10)$ denote an accuracy parameter. 
\item Let $\delta \in (0,1/10)$ denote a failure probability.
\item Let $\Pi\in \R^{s\times n^2}$ be a {\sf TensorSRHT} matrix.
\item Let  $s=\Theta(\epsilon^{-2} m \log^3(n m / (\epsilon\delta) ))$.
\end{itemize}
Then we have  
\begin{align*}
   \Pr[  \|\Pi{\sf B}^\top x\|_2 = & ~ (1\pm\epsilon) \|{\sf B}^\top x\|_2, \forall x \in \R^m ] \geq 1-\delta.
\end{align*}

Moreover, there is a single pass streaming algorithm that uses $O(s^2 + n^2)$ space and can store $\Pi{\sf B}^\top$ by only reading constraints $A_1, \cdots, A_m$ once. The matrix $\Pi {\sf B}^\top$ can be computed in time 
\begin{align*}
    O(m\cdot \Tmat(s,n,n)).
\end{align*}

\end{lemma}

\begin{proof}
The correctness part follows directly from Lemma~\ref{lem:tensor_srht_implies_ose}. It remains to argue for the running time. We need to unravel the construction of both $\Pi$ and ${\sf B}$. 

Recall that 
\begin{align*}
\Pi=\frac{1}{\sqrt s}P\cdot (HD_1\otimes HD_2)
\end{align*}
and 
\begin{align*}
    & ~ \Pi(S(x)^{-1/2}\otimes S(x)^{-1/2})\vect(A_i) \\
    = & ~ \frac{1}{\sqrt s} P\cdot (HD_1\otimes HD_2)\cdot (S(x)^{-1/2}\otimes S(x)^{-1/2})\vect(A_i) \\
    = & ~ \frac{1}{\sqrt s} P\cdot (HD_1S(x)^{-1/2}\otimes HD_2S(x)^{-1/2})\vect(A_i) \\
    = & ~ \frac{1}{\sqrt s}P\cdot \vect(HD_2S(x)^{-1/2}A_iS(x)^{-1/2}D_1H^\top). 
\end{align*}
Since $P$ is a row sampling matrix, the product can be computed as follows:

{\bf First Step.}
 First compute $HD_1S(x)^{-1/2}$ and $HD_2S(x)^{-1/2}$. Since $H$ is a Hadamard matrix, this step can be carried out in $O(n^2\log n)$ time. To store them, we only need $O(n^2)$ space.

{\bf Second Step.}
    Applying $P$ to the vector can be interpreted as sampling $s$ coordinates from the matrix $HD_2S(x)^{-1/2}A_iS(x)^{-1/2}D_1H^\top$. Let $(i_1,j_1),\ldots,(i_s,j_s)$ denote the coordinates sampled by $P$. We construct two matrices $X, Y\in \R^{s\times n}$ such that the $k$-th row of $X$ is $(HD_2S(x)^{-1/2})_{i_k,*}$ and the $k$-th row of $Y$ is $(HD_1S(x)^{-1/2})_{j_k,*}$. It is easy to verify that the $(i_k,j_k)$-th entry of $XA_iY^\top \in \R^{s \times s}$ is the corresponding entry of $HD_2S(x)^{-1/2}A_iS(x)^{-1/2}D_1H^\top$. This step therefore takes $\Tmat(n, n, s)$ time and storing $XA_i Y^\top$ takes $O(s^2)$ space. After the sampling process, we end up with a vector with $s$ entries, which means that we only need spend $O(sm)$ space over all $m$ columns. As $s\geq m$, the $O(sm)$ space is subsumed by $O(s^2)$. For runtime, we need to repeat this procedure for all $m$ columns, resulting in a total runtime of $O(m\cdot \Tmat(s,n,n))$.
\end{proof}

\section*{Acknowledgement}

Lichen Zhang is supported by NSF grant No.\ 1955217 and No.\ 2022448.

\ifdefined\isarxiv
\bibliographystyle{alpha}
\bibliography{ref}
\else
\bibliography{ref}
\bibliographystyle{alpha}

\fi

\newpage
\onecolumn
\appendix





\end{document}